\documentclass[journal,draftcls,onecolumn,12pt,twoside]{IEEEtranTCOM}
\ifCLASSINFOpdf
  \usepackage[pdftex]{graphicx}
\else
  \usepackage[dvips]{graphicx}
\fi

\usepackage{epsfig}
\usepackage{color}
\usepackage{amssymb}
\usepackage{amstext}
\usepackage{amsmath}
\usepackage{amsthm}
\usepackage{multicol}
\usepackage{pslatex}
\usepackage{float}
\usepackage[small]{caption}
\usepackage{subcaption}
\usepackage[update,prepend]{epstopdf}
\usepackage{lmodern}
\usepackage{cite}
\usepackage{balance}
\usepackage{listings}
\def\postbreak{\raisebox{0ex}[0ex][0ex]{\ensuremath{\hookrightarrow\space}}}
\begin{document}

\title{Comment on ``Analysis of a Charge-Pump PLL: A New Model'' by M. van Paemel}

\author{
  Kuznetsov~N.V., Yuldashev~M.V., Yuldashev~R.V., 
  Blagov~M.V., Kudryashova~E.V., Kuznetsova~O.A., Mokaev~T.N.
\thanks{
}
\thanks{
N.V. Kuznetsov$^{\,a,b,c}$, M.V. Yuldashev$^{\,a}$,
R.V. Yuldashev$^{\,a}$, M.V. Blagov$^{\,a,b}$, E.V. Kudryashova$^{\,a}$, 
O.A. Kuznetsova$^{\,a}$, and T.N. Mokaev$^{\,a}$ are from
($^a$) Faculty of Mathematics and Mechanics,
Saint-Petersburg State University, Russia;
($^b$) Dept. of Mathematical Information Technology,
University of Jyv\"{a}skyl\"{a}, Finland;
($^c$) Institute of Problems of Mechanical Engineering RAS, Russia;
(corresponding author email: nkuznetsov239@gmail.com).
}
}

\maketitle

\begin{abstract}

  In this short communication we comment on the
non-linear mathematical model of CP-PLL introduced by V.Paemel.
We reveal and obviate some shortcomings in the model.
\end{abstract}

\IEEEpeerreviewmaketitle

\section{Introduction}
M. van Paemel's article \cite{Paemel-1994} was the first one where a complete nonlinear mathematical model of CP-PLL is derived.
The classical models (see e.g. \cite{Gardner-2005-book,Best-2007,LeonovKYY-2015-TCAS}) considered approximation of the phase detector dynamics in continuous time and linearization.
While approximate models are useful for analysis of small frequency deviations
between VCO and Ref signals, Paemel's models is exact and can also be used for studying out-of-lock behavior.

However, the algorithm suggested in \cite{Paemel-1994} does not always work.
Below we reveal and obviate shortcomings in the Paemel's model and discuss corresponding numerical examples.


\section{Numerical examples}
\label{sec:num-examples}
The following examples demonstrate that algorithm and formulas suggested by M. van Paemel
should be used carefully for simulation even inside allowed area (see original paper Fig~18 and Fig.~22).
While the examples are given for the first time, the main idea of Example 1 was already noticed by P. Acco and O. Feely \cite{acco2003etude,Orla-2012}. 
P. Acco and O. Feely considered only near-locked state, therefore they didn't notice problems with out-of-lock behavior. Example 2 and Example 3 demonstrate problems with out-of-lock behavior, which was not discovered before.

\subsection{Example 1}
Consider the following set of parameters and initial state:
\begin{equation}
\label{ex1}
\begin{aligned}
& R_2 = 0.2;
C = 0.01;
K_v = 20;
I_p = 0.1;
T = 0.125;\\
& \tau(0) = 0.0125;
v(0) = 1.
\end{aligned}
\end{equation}
Calculation of normalized parameters (equations (27)-(28) and (44)-(45) in \cite{Paemel-1994})
\begin{equation}
\label{F_n zeta}
\begin{aligned}
  & K_N = I_pR_2K_vT = 0.05,\\
  & \tau_{2N}=\frac{R_2C}{T} = 0.016,\\
  & F_N = \frac{1}{2\pi}\sqrt{\frac{K_N}{\tau_{2N}}} \approx 0.2813,\\
  & \zeta = \frac{\sqrt{K_N\tau_{2N}}}{2} \approx 0.0141,
\end{aligned}
\end{equation}
shows that parameters \eqref{ex1} correspond to allowed area in Fig.~\ref{fig:allowed-domain} (equations (46)--(47), Fig~18 and Fig.~22 in \cite{Paemel-1994}):
\begin{figure}[H]
  \centering
  \includegraphics[width=1\linewidth]{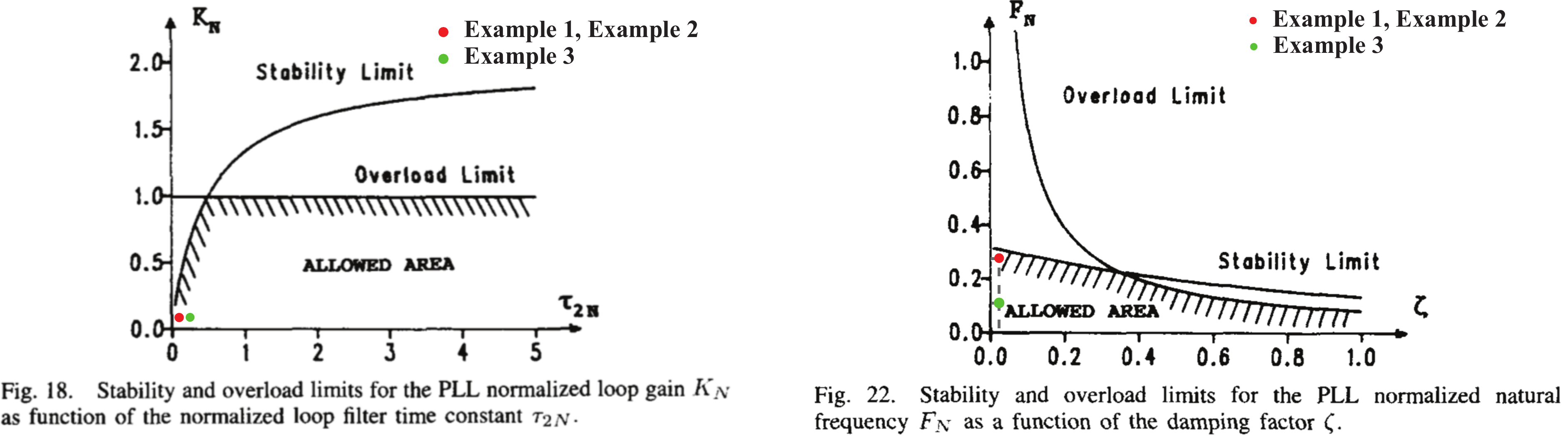}
  \caption{Parameters for Example 1, Example 2, and Example 3 correspond to allowed area (see Fig~18 and Fig.~22 in \cite{Paemel-1994})}
  \label{fig:allowed-domain}
\end{figure}
\begin{equation}
\label{allowed domain eq}
  \begin{aligned}
    & F_N < \frac{\sqrt{1+\zeta^2}-\zeta}{\pi} \approx 0.3138,
    \\
    & F_N < \frac{1}{4\pi\zeta} \approx 5.6438. 
  \end{aligned}
\end{equation}
\begin{figure}[h]
  \centering
  \includegraphics[width=0.9\linewidth]{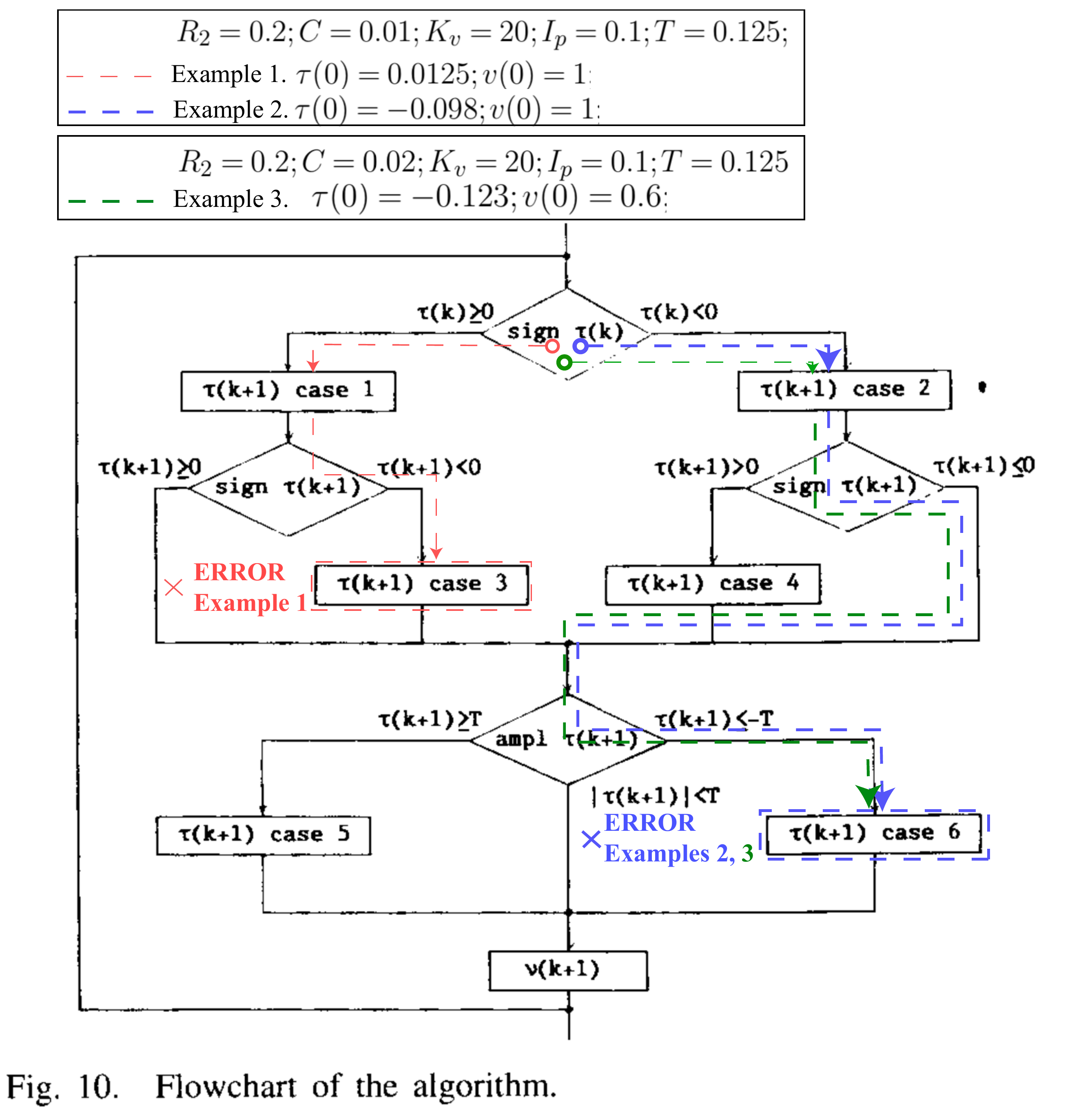}
  \caption{Demonstration of Example 1, Example 2, and Example 3 in the flowchart of the algorithm (see Fig.~10 in \cite{Paemel-1994})}
  \label{fig:paemel-flowchart}
\end{figure}
Now we use the flowchart in Fig.~\ref{fig:paemel-flowchart} (Fig.~10 in \cite{Paemel-1994}) to compute $\tau(1)$ and $v(1)$: since $\tau(0) > 0$ and $\tau(0) < T$,
we proceed to \emph{case 1)} and corresponding relation for $\tau(k+1)$ (equation (7) in \cite{Paemel-1994}):
\begin{equation}
  \label{paemel-case-1}
  \begin{aligned}
    & \tau(k+1)= \frac{-I_pR_2-v(k)+\sqrt{(I_p R_2 + v(k))^2 - \frac{2I_p}{C}(v(k)(T - \tau(k)) - \frac{1}{K_v})}}{\frac{I_p}{C}}.
  \end{aligned}
\end{equation}
However, the expression under the square root in \eqref{paemel-case-1} is negative:
\begin{equation}
\begin{aligned}
  & (I_p R_2 + v(0))^2 - \frac{2I_p}{C}(v(0)(T - \tau(0)) - \frac{1}{K_v}) = -0.2096 < 0.
\end{aligned}
\end{equation}
Therefore the algorithm is terminated with error. 

\subsection{Example 2}
Consider the same parameters as in Example 1, but $\tau(0) = -0.098$:
\begin{equation}
\label{ex2}
\begin{aligned}
&
R_2 = 0.2;
C = 0.01;
K_v = 20;
I_p = 0.1;
T = 0.125;\\
& \tau(0) = -0.098;
\quad v(0) = 1.
\end{aligned}
\end{equation}
In this case \eqref{F_n zeta}, \eqref{allowed domain eq}, and Fig.~\ref{fig:allowed-domain} are the same as in Example 1, i.e. we are in the ``allowed area''.
Now we compute $\tau(1)$ and $v(1)$  following the flowchart in Fig.~\ref{fig:paemel-flowchart}: since $\tau(0) < 0$
we proceed to \emph{case 2)} and corresponding equation of $\tau(k+1)$ (equation (9) in \cite{Paemel-1994}):
\begin{equation}
  \label{paemel-case-2}
  \begin{aligned}
    & \tau(1) 
      = \frac{\frac{1}{K_v}-I_pR_2\tau(0)-\frac{I_p\tau(0)^2}{2C}}{v(0)}
        -T+\tau(0) 
      = -0.21906,\\
    & -0.2191 < -T = -0.125.
  \end{aligned}
\end{equation}
This fact indicates cycle-slipping (out of lock).
According to the flowchart in Fig.~\ref{fig:paemel-flowchart} (see Fig.~10 in \cite{Paemel-1994}), we should proceed to \emph{case 6)}
and recalculate $\tau(1)$.
First step of case 6) is to calculate $t_1, t_2, t_3,...$ (equations (16) and (17) in \cite{Paemel-1994}):
\begin{equation}
  \label{case-6-n}
  \begin{aligned}
    & t_n = \frac{v_{n-1}-I_pR_2-\sqrt{(v_{n-1}-I_pR_2)^2-2\frac{I_p}{C}\cdot \frac{1}{K_v}}}{\frac{I_p}{C}},\\
    &v_n = v_{n-1}-\frac{I_p}{C}t_n,\\
    &v_0 = v(k-1).
  \end{aligned}
\end{equation}
Since $k = 0$, then
\begin{equation}
  \begin{aligned}
    & t_1 = \frac{v_{0}-I_pR_2-\sqrt{(v_{0}-I_pR_2)^2-2\frac{I_p}{C}\cdot \frac{1}{K_v}}}{\frac{I_p}{C}},\\
    &v_1 = v_{0}-\frac{I_p}{C}t_1,\\
    &v_0 = v(-1).
  \end{aligned}
\end{equation}
However, $v(-1)$ doesn't make sense and algorithm terminates with error.\footnote{However, this can be fixed, by setting $v(-1)=v(0)-\frac{I_p}{C}\tau(0)$.}
Even if we suppose that it is a typo and $v_0 = v(0)$, then relation under the square root become negative:
\begin{equation}
\begin{aligned}
  & (v(0) - I_p R_2)^2 - 2\frac{I_p}{CK_v} = -0.0396 < 0.
\end{aligned}
\end{equation}
In both cases the algorithm is terminated with error. 
Note, that modification of case 2) corresponding to VCO overload 
(equation (35) in \cite{Paemel-1994}) can not be applied here, 
since $v(0) > I_p R_2$ (no overload) and $v(1)$ is not computed yet because of the error.

\subsection{Example 3}
Consider parameters:
\begin{equation}
\label{ex3}
\begin{aligned}
& \tau(0) = -0.123;
\quad v(0) = 0.6,\\
&
R_2 = 0.2;
C = 0.02;
K_v = 20;
I_p = 0.1;
T = 0.125.
\end{aligned}
\end{equation}
Similar to \eqref{F_n zeta} and \eqref{allowed domain eq}
\begin{equation}
\begin{aligned}
  & K_N = 0.05,
   \quad \tau_{2N} = 0.032,
  \\
  & F_N \approx 0.1989,
    \quad  \zeta = 0.02,
\end{aligned}
\end{equation}
\begin{equation}
\label{allowed domain eq3}
  \begin{aligned}
    & F_N < \frac{\sqrt{1+\zeta^2}-\zeta}{\pi} \approx 0.3120,
    \\
    & F_N < \frac{1}{4\pi\zeta} \approx 3.9789,
  \end{aligned}
\end{equation}
parameters \eqref{ex3} correspond to allowed area in Fig.~\ref{fig:allowed-domain} (equations (46)-(47), Fig.~18 and Fig.~22 in \cite{Paemel-1994}).

Now we compute $\tau(1)$ and $v(1)$  following the flowchart in Fig.~\ref{fig:paemel-flowchart}: since $\tau(0) < 0$
one proceeds to \emph{case 2)} and corresponding equation for computing $\tau(k+1)$ (equation (9) in \cite{Paemel-1994}):
\begin{equation}
\label{paemel-case-2}
  \begin{aligned}
    & \tau(1) = \frac{\frac{1}{K_v}-I_pR_2\tau(0)-\frac{I_p\tau(0)^2}{2C}}{v(0)}
      -T+\tau(0)
      \approx -0.224,\\
    & -0.224 < -T = -0.125.
  \end{aligned}
\end{equation}
The last inequality indicates cycle-slipping (out of lock).
According to the flowchart in Fig.~\ref{fig:paemel-flowchart} (see Fig.~10 in \cite{Paemel-1994}), one proceeds to \emph{case 6)}
and recalculates $\tau(1)$.
First step of \emph{case 6)} is to calculate $t_1, t_2, t_3,...$ using \eqref{case-6-n} (see equations (16) and (17) in \cite{Paemel-1994}) until $t_1+t_2+\ldots+t_n>|\tau(0)|$.
Even if we suppose $v(-1)=v(0)-\frac{I_p}{C}\tau(0)$, we get
\begin{equation}
  \begin{aligned}
    & t_1 = 0.0463,\ v_1 = 1.215;\\
    & t_2 = 0.0618,\ v_2 = 0.983;\\
    & t_1+t_2=0.1081<|\tau(0)|=0.123.
  \end{aligned}
\end{equation}
However, $t_3$ can not be computed, because the relation under the square root in \eqref{case-6-n} is negative:
\begin{equation}
  \begin{aligned}
    & (v_2-I_pR_2)^2-2\frac{I_p}{C}\cdot \frac{1}{K_v} \approx -0.0726.
  \end{aligned}
\end{equation}

\section{Corrected discrete time model of CP-PLL}
Below we suggest the corrected \emph{discrete time nonlinear mathematical model of CP-PLL},
in which shortcomings are fixed.
The problem with flowchart (see Fig.~\ref{fig:paemel-flowchart}) is that
the sign of $\tau(k+1)$ computed by \emph{case 1)} is used to decide  whether \emph{case 3)} should be used or not.
Similarly, \emph{case 2)} always precede \emph{cases 4),5),6)} which may lead to errors.
However, it is possible to use $\tau(k)$ and $v(k)$ explicitly to select correct formula for $\tau(k+1)$.
This allows one to avoid computing square roots of negative numbers, reduce number of cases from 6 to 4, and apply methods from theory of discrete time dynamical systems (see, e.g. \cite{Orla-2012}).

Here $v(0)$ and $\tau(0)$ are initial conditions (Paemel's notation is used).
\begin{equation}
\label{complete-model}
  \begin{aligned}
    & \tau(k+1) =
    \left\{\begin{array}{l}
      \frac{-b + \sqrt{b^2 - 4ac}}{2a},
      \ \quad \tau(k) \geq 0, \quad c \leq 0,
      \\
      \\
      \frac{1}{\omega_{\rm vco}^{\text{free}} + K_vv(k)}
- T + (\tau(k) \text{ mod } T),
      \\
      \quad\quad\quad\quad\quad\quad\quad
       \tau(k) \geq 0, \quad c > 0,
      \\
      \\
      l_b-T,
      \quad\quad\quad
       \tau(k) < 0, \quad l_b \leq T,
      \\
      \\
      \frac{-b + \sqrt{b^2 - 4ad}}{2a},
      \quad \tau(k) < 0, \quad l_b > T,
    \end{array}
    \right.\\
    & v(k+1) = v(k)+\frac{I_p}{C}\tau(k+1).\\
          \end{aligned}
    \end{equation}
    \begin{equation}
    \nonumber
      \begin{aligned}
    & a = \frac{K_vI_p}{2C},
    \\
    &
      b = b(v(k)) = \omega_{\rm vco}^{\text{free}} + K_vv(k)
        + K_vI_pR_2,
    \\
    &
      c = c(\tau(k),v(k)) =
     (T - (\tau(k)\ {\rm mod}\ T))
\left(
  \omega_{\rm vco}^{\text{free}} + K_vv(k)
\right) - 1,\\
    & l_b = l_b(\tau(k),v(k)) =\tfrac{1 - S_{l_a}}{K_vv(k)+\omega_{\rm vco}^{\text{free}}},\\
    & S_{l_a} = S_{l_a}(\tau(k),v(k)) = S_{l_k}\text{ mod }1,\\
    & S_{l_k} = S_{l_k}(\tau(k),v(k)) =
      -\left(
        K_vv(k) - I_p R_2 K_v +\omega_{\rm vco}^{\text{free}}
      \right)\tau(k)
      + K_vI_p\tfrac{\tau(k)^2}{2C},
    \\
    & d = d(v(k)) = S_{l_a}+T(K_vv(k)+\omega_{\rm vco}^{\text{free}}) - 1.
  \end{aligned}
\end{equation}
Here VCO frequency is $f_{\rm vco}=\omega_{\rm vco}^{\text{free}}+K_v v_c$, and $\omega_{\rm vco}^{\text{free}}$ is a free-running (quiescent) frequency (in V.Paemel's paper $\omega_{\rm vco}^{\text{free}}=0$).
If at some point VCO becomes overloaded
 one should stop simulation or use another set of equations, based on ideas similar to (34) and (35) in \cite{Paemel-1994}.
 Overload conditions are
 \begin{equation}
\label{overload eq}
  \begin{aligned}
    &\tau(k)>0 \text{ and }
    v(k)+\frac{\omega_{\rm vco}^{\rm free}}{K_{\rm vco}}-\frac{I_p}{C}\tau_k<0,\\
    & \tau(k)<0 \text{ and }
    v(k)+\frac{\omega_{\rm vco}^{\rm free}}{K_{\rm vco}}-I_p R_2<0.\\
  \end{aligned}
\end{equation}

Remark that following the ideas from \cite{Paemel-1994,acco2003etude,Orla-2013-review},
the number of parameters in \eqref{complete-model} can be reduced
to just two ($\alpha$ and $\beta$) by the following change of variables
\begin{equation}
\label{alpha-beta-eqs}
  \begin{aligned}
    & s(k) = \frac{\tau(k)}{T},\
     \omega(k)=T
      \left(
        \omega_{\rm vco}^{\text{free}} + K_vv(k)
      \right) - 1,\\
    & \alpha = K_vI_pTR_2,\
     \beta = \frac{K_vI_pT^2}{2C}.
  \end{aligned}
\end{equation}

\section{Numerical examples for corrected model}
Consider application of the corrected model \eqref{complete-model} to numerical examples from section~\ref{sec:num-examples}.
All three examples assume $\omega_{\rm vco}^{\rm free} = 0$.
\subsection{Example 1}
\begin{figure}[H]
  \centering
  \includegraphics[width=0.5\linewidth]{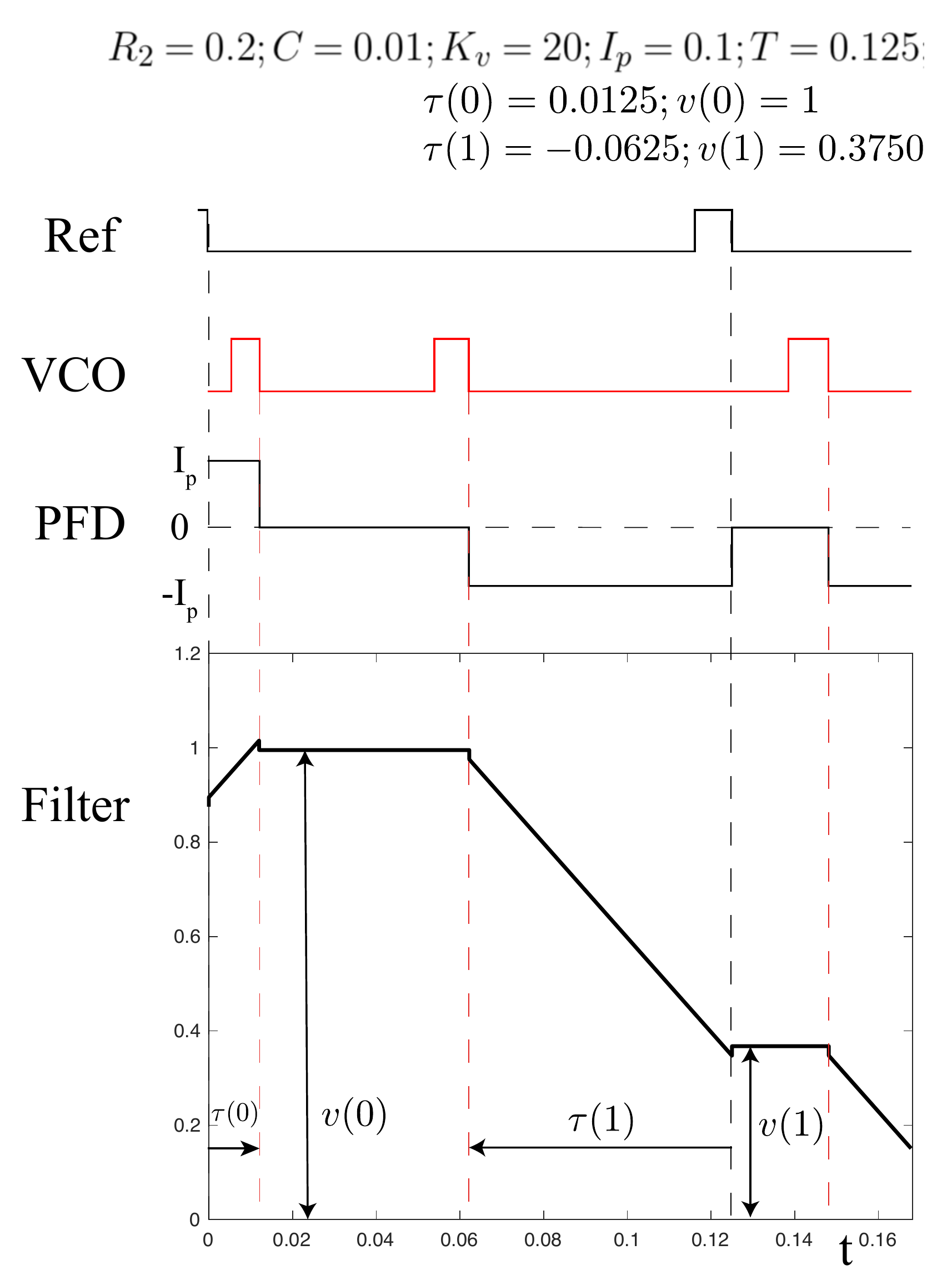}
  \caption{Application of corrected model to Example 1: Reference signal, VCO output, PFD output, and filter output.}
  \label{fig:example-1-vco-input}
\end{figure}

By \eqref{complete-model} and \eqref{ex1} we calculate value of $c$:
\begin{equation}
\begin{aligned}
  & c =
     (T - (\tau(0)\ {\rm mod}\ T))
     K_v v(0) - 1 = 1.2500\\
\end{aligned}
\end{equation}
and since $\tau(0) \geq 0$ and $c > 0$ we get
\begin{equation}
\begin{aligned}
  & \tau(1) = \frac{1}{K_v v(0)}
- T + (\tau(0) \text{ mod } T) = -0.0625,
  \\
  & v(1) = v(0)+\frac{I_p}{C}\tau(1) = 0.3750.
\end{aligned}
\end{equation}
Illustration of this example is shown in Fig.~\ref{fig:example-1-vco-input}. 

Note, that in this case there is no VCO overload (no saturation), since the filter output (VCO input) is positive, see Fig.~\ref{fig:example-1-vco-input}.

\subsection{Example 2}
\begin{figure}[H]
  \centering
  \includegraphics[width=0.5\linewidth]{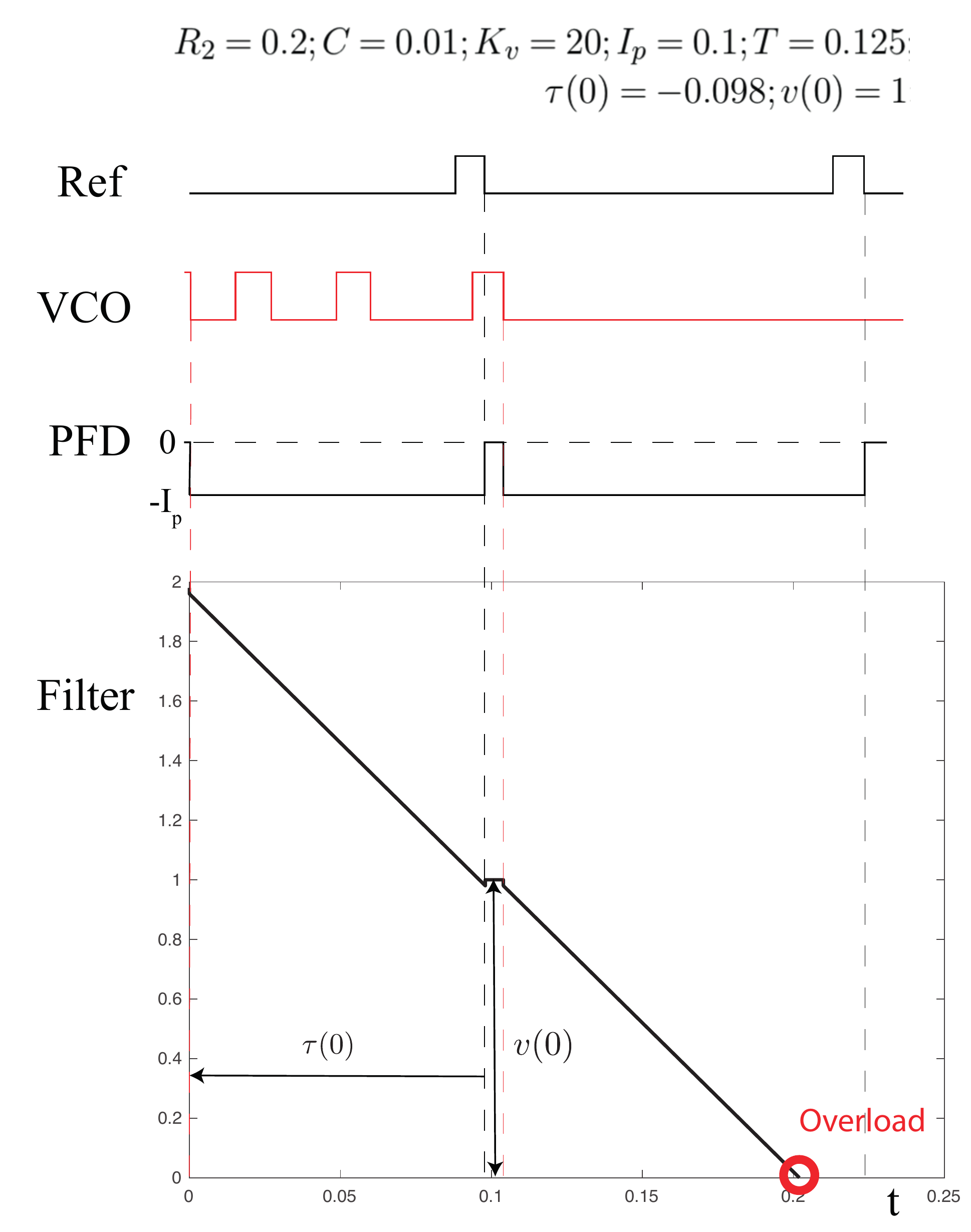}
  \caption{Application of corrected model to Example 2: Reference signal, VCO output, PFD output, and filter output.}
  \label{fig:example-2-vco-input}
\end{figure}

By \eqref{complete-model} and \eqref{ex2} we have $l_b \approx 0.0059$.
Since $l_b \leq T$, then $\tau(1) \approx -0.1191, v(1) \approx -0.1906$.
In this case the VCO is overloaded (see Fig.~\ref{fig:example-2-vco-input}). 
Model \eqref{complete-model} correctly detects overload
by \eqref{overload eq}
\begin{equation}
 \begin{aligned}
   & v(1)+\frac{\omega_{\rm vco}^{\rm free}}{K_{\rm vco}}-I_p R_2 \approx -0.2106 < 0
   \\
 \end{aligned}
 \end{equation}
  and stops simulation.

\subsection{Example 3}
\begin{figure}[H]
  \centering
  \includegraphics[width=0.5\linewidth]{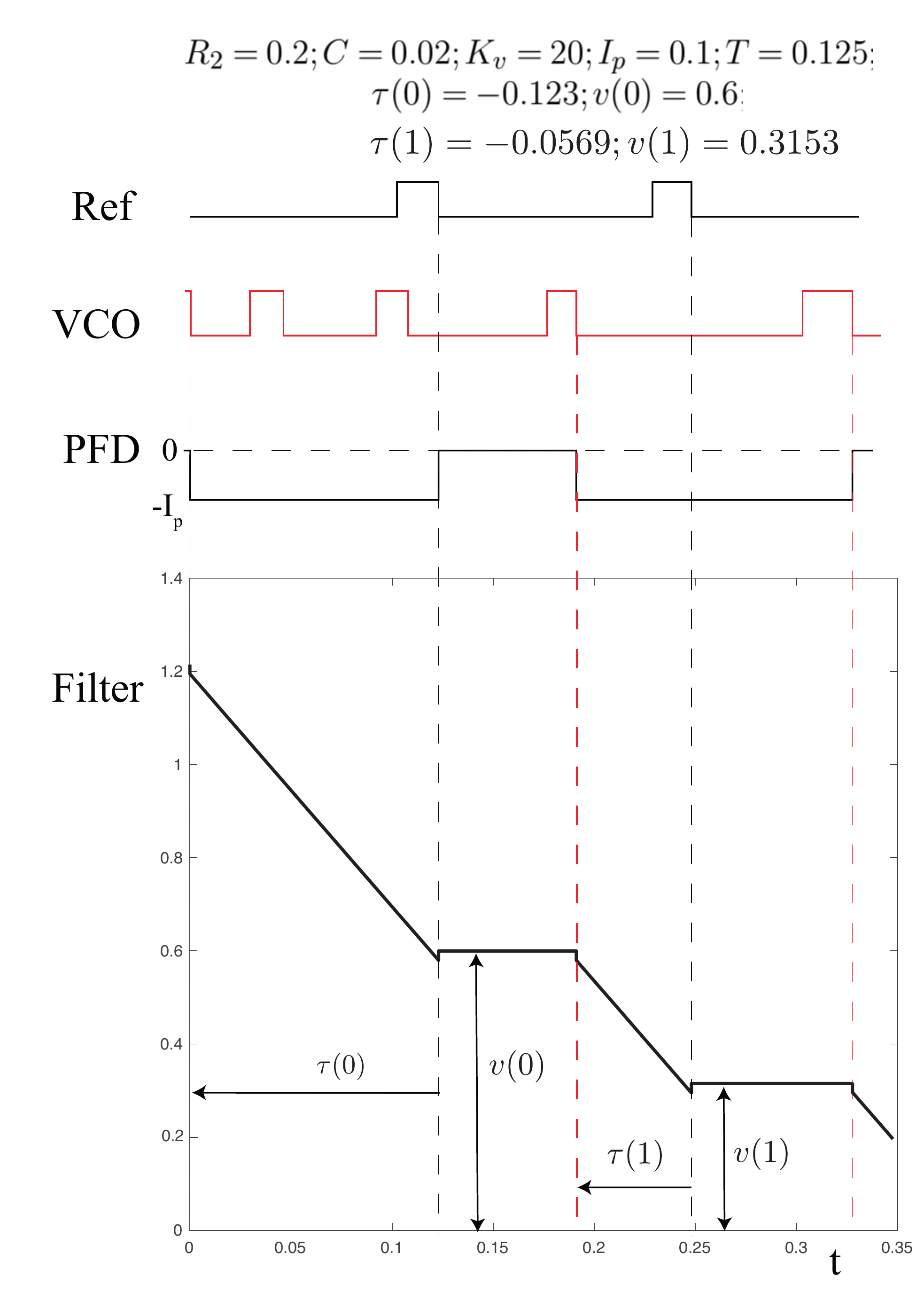}
  \caption{Application of corrected model to Example 3: Reference signal, VCO output, PFD output, and filter output.}
  \label{fig:example-3-vco-input}
\end{figure}

Note, that in this case VCO is not overloaded, since the filter output (VCO input) is positive, see Fig.~\ref{fig:example-3-vco-input}.
Equations \eqref{complete-model} allow to correctly calculate next step:
\begin{equation}
\begin{aligned}
  & \tau(1) = 0,
  \quad v(1) = 10.
  \\
\end{aligned}
\end{equation}

\section{Comparison of Simulink vs V.Paemel's model vs Corrected model}
\begin{figure}[H]
  \centering
  \includegraphics[width=1\linewidth]{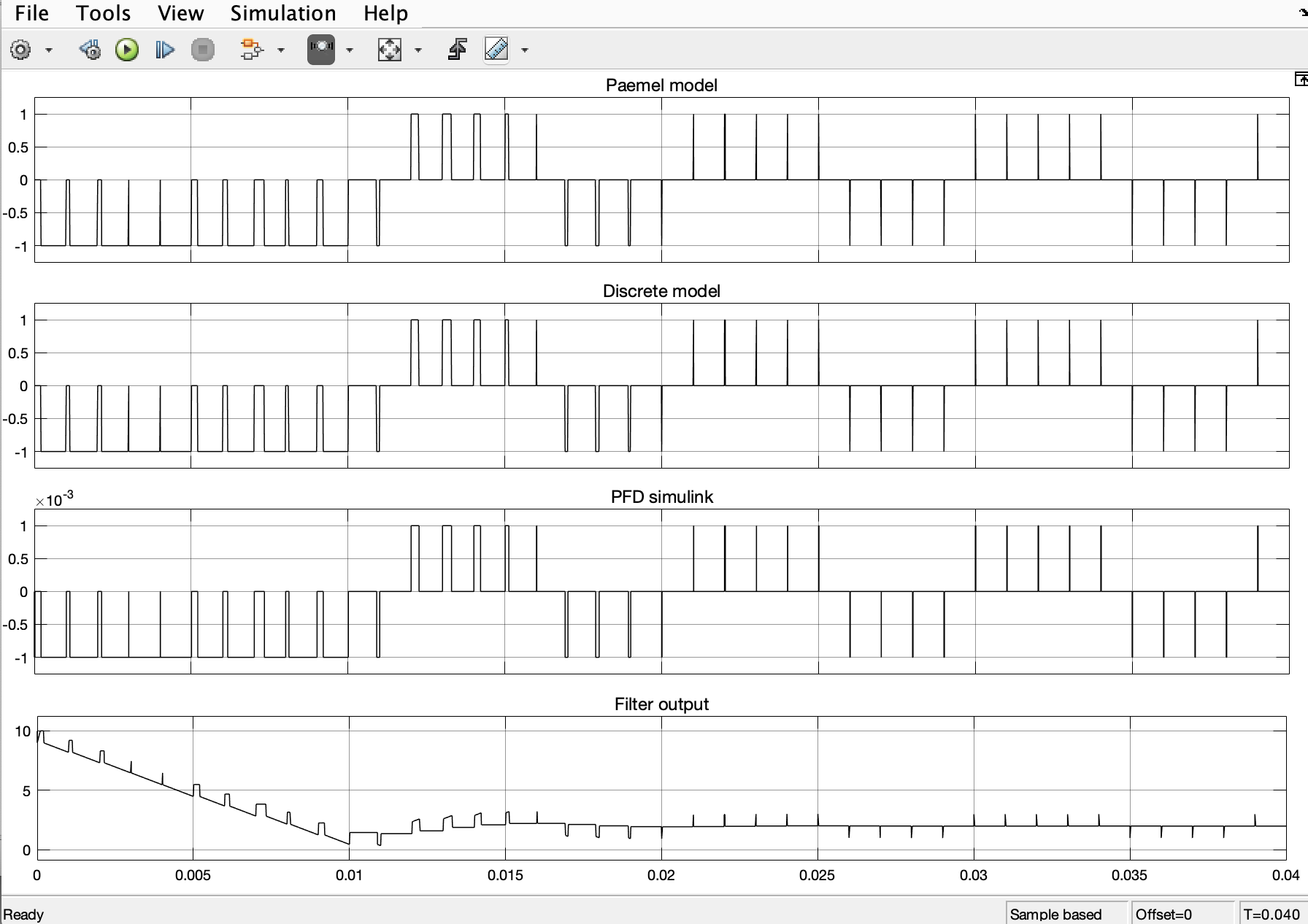}
  \caption{Comparison of PFD outputs of Simulink model (PFD Simulink) vs V.Paemel's model (Paemel model) vs Corrected model (Discrete model).
  Lower subfigure demonstrates output of Loop filter.
  For considered set of parameters 
  ($\tau(0) = 0;
  v(0) = 10;
  R_2 = 1000;
  C = 10^{-6};
  K_v = 500;
  I_p = 10^{-3};
  T = 10^{-3};
  \tau_{2N} = 1;
  K_N = 0.5;
  F_N = 0.1125;
  \zeta =0.3536$)
  all three models produce almost the same results.}
 \label{fig:sim vs paemel vs correcter}
\end{figure}

Correctness of proposed model was verified by extensive simulation
in Matlab Simulink.
Circuit level model in Matlab Simulink was compared with original
model by V.~Paemel and proposed model (see parameters \eqref{Simulink parameters} and Fig.~\ref{fig:sim vs paemel vs correcter}).
\begin{equation}
\label{Simulink parameters}
\begin{aligned}
& \tau(0) = 0;
\quad v(0) = 10,\\
&
R_2 = 1000;
C = 10^{-6};
K_v = 500;
I_p = 10^{-3};
T = 10^{-3}.
\\
& \tau_{2N} = 1;
K_N = 0.5;
F_N = 0.1125;
\zeta =0.3536.
\end{aligned}
\end{equation}
Based on simulation for this set of parameters
all three models produce almost the same results.

\section{Conclusions}
There were many attempts to generalize equations derived in \cite{Paemel-1994} for higher-order loops (see, e.g.\cite{hanumolu2004analysis,Shakhtarin2014-pfd,bi2011linearized,hangmann2014stability,milicevic2008time,sancho2004general}), but the resulting transcendental equations can not be solved analytically without using approximations.

\section*{Acknowledgments} \label{sec:acknowledgement}
The work is supported by the Russian Science Foundation project 19-41-02002.

\balance
\bibliographystyle{IEEEtranTCOM}

\section*{Illustrations for numerical examples}

\end{document}